# Beyond Human-Likeness: Mapping the Scientific Critique Profiles of LLMs and Human Reviewers

Yunhan Yang[1], Mike Thelwall[1*], Guoxiu He[2]

[1] School of Information, Journalism and Communication, The University of Sheffield, Sheffield, UK

[2] School of Economics and Management, East China Normal University, Shanghai, China

**Abstract**

Large language models (LLMs) are increasingly discussed as tools for peer review, but their value is often assessed through human-likeness, perceived usefulness, or textual overlap with reviewer comments. This study shifts attention from whether LLMs resemble human reviewers to what functions of scientific critique they perform. Using ICLR 2025 peer-review data, we compare human reviews with LLM reviews generated under baseline and expert prompts. We operationalize scientific critique through two review acts, weakness critique and scientific questioning, and annotate point-level review text using five theory-guided frameworks: Anderson's knowledge types, Toulmin's argumentation model, Graesser's question depth, SOLO cognitive complexity, and Hattie's feedback functions. The results reveal a differentiated critique profile. Human reviews placed greater emphasis on scientific framing and revision guidance, more often identifying higher-order weaknesses and asking questions oriented toward improvement. LLM reviews showed higher rates of explanatory depth, integrative reasoning, and explicit argument structuring. Expert prompting did not make LLM critique uniformly more human-like; it partially narrowed some gaps but mainly amplified LLM-specific tendencies toward integration and formal argumentation. These findings show that LLM-assisted peer review changes the functional composition of review text, making it important to distinguish LLM-amplified critique from areas requiring human prioritization and accountable judgement.

## 1 Introduction

Large language models (LLMs) are increasingly becoming part of scholarly communication and research evaluation. They are discussed not only as tools for writing, summarizing, or retrieving scientific texts, but also as systems that may help evaluate research quality, generate manuscript feedback, and support peer review workflows (Kousha & Thelwall, 2025; Lund et al., 2023; Thelwall, 2025a). Their entry into peer review intersects with longstanding pressures in scholarly publishing, including reviewer scarcity, delays, uneven feedback quality, and concerns about bias and accountability (Mahony, 2022). These pressures make peer review a natural site for LLM support, but also a critical setting for asking what kinds of evaluative work LLMs perform when they enter scholarly evaluation.

Recent work shows that LLMs can already participate in peer-review-like tasks in practically consequential ways. LLM-generated feedback can overlap with human reviewer comments, may be perceived as useful, specific, or actionable, and can influence how reviewers revise their own reviews in real peer-review workflows (Liang, Zhang, et al., 2024; Thakkar et al., 2026). Studies of AI-modified reviews in machine-learning conferences further suggest that LLMs are already entering real review text, raising questions about how they may change review style and evaluative practice (Liang, Izzo, et al., 2024). Benchmark studies using ICLR or related machine-learning review data also show uneven performance: LLMs can perform well on some review-related tasks, but remain limited in reliable review-revision reasoning, critical feedback, novelty assessment, and balanced multidimensional evaluation (Loc et al., 2026; K. Wu et al., 2025; Xu et al., 2024; Zhou et al., 2024).

Despite the above, it is not known how different scientific critique functions are represented in human and LLM-generation texts, and whether LLMs are weak relative to humans for any important review functions. A report can be fluent, detailed, and well-structured without identifying the most scientifically consequential weakness; conversely, a short reviewer question may reflect deep field-specific judgment. Recent work on peer-review quality similarly emphasizes that useful reviews are multidimensional, involving relevance, specificity, constructiveness, thoroughness, and usefulness for authors and editors (Sizo et al., 2026). Evaluating LLM reviews therefore requires attention to which critique functions are reproduced, amplified, under-represented, or reconfigured in LLM-generated reviews.

This functional view also connects to broader debates about when AI makes a difference in science. If AI is especially useful in fragmented or combinatorially complex knowledge spaces, where progress depends on connecting dispersed ideas, methods, and evidence (Bianchini et al., 2026), then LLM-generated reviews may be especially likely to display mechanistic questions or integrative critique. Yet recombining knowledge elements is not the same as making expert judgments about scientific significance, evidential adequacy, or revision priorities. Understanding LLMs in peer review therefore requires a profile of their critique functions, rather than a single judgment of whether their reviews are human-like.

We used the structure of the International Conference on Learning Representations (ICLR) review form as the empirical entry point for this analysis. The distinction between weakness critique and reviewer questions followed the review form, which asks reviewers to report weak points and questions as separate sections. These two sections make different aspects of scientific critique observable. Weakness sections capture critique as diagnosis: what reviewers identify as problematic and how they justify that judgment. Question sections capture critique as inquiry: what reviewers ask authors to clarify, explain, connect, or improve. Treating these sections separately allowed us to compare the content and argumentation of criticism with the depth, integration, and feedback function of scientific questioning.

We also examined whether these patterns are sensitive to prompt specification. The baseline condition used an ICLR-style review prompt that asks the model to produce weak points and questions following standard reviewer instructions. The expert-prompt condition preserved the same output structure but placed greater emphasis on concerns commonly associated with expert human review, such as problem importance, claim-method-evidence alignment, robustness, and contribution. This contrast allowed us to test whether observed human-LLM differences mainly reflect ordinary prompt specification or a more stable difference in functional profile of LLM critique.

The study addresses three research questions:

RQ1. How does baseline LLM critique differ from human critique in the diagnostic content and argumentative support of weaknesses?

RQ2. How does baseline LLM questioning differ from human questioning in explanatory depth, integrative complexity, and revision-oriented feedback function?

RQ3. Does expert prompting move LLM critique closer to human critique, or amplify a distinct LLM critique profile?

Empirically, we examine ICLR 2025 peer review using a large-scale corpus containing more than 11,000 submissions. The main comparison is conducted on matched human and LLM review data, and the expert-prompt analysis uses a stratified sample of 600 papers across accepted, rejected, and withdrawn submissions. We decompose review text into point-level weakness and question units and annotate them using five theory-guided frameworks. For weaknesses, these frameworks capture the content target and argumentative structure of critique; for questions, they capture explanatory depth, integrative complexity, and feedback function. The annotations are then aggregated into paper-level metrics for comparing human, baseline LLM, and expert-prompt LLM critique.

This article makes three contributions. First, it reframes LLM peer review as a question of evaluative function rather than overall human-likeness or replacement. Second, it develops a theory-guided approach for treating peer review reports as structured evaluative information, translating weakness and question points into interpretable paper-level metrics. Third, it shows that human and LLM-generated reviews differ systematically in the critique functions they emphasize, providing a diagnostic basis for assessing how LLMs may reshape peer review without assuming that any type of scientific judgment can be safely delegated to them.

## 2 Literature Review

This review develops the basis for comparing human and LLM-generated peer reviews as functional critique profiles rather than as outputs on a single quality scale. Section 2.1 reviews evidence on LLMs in scholarly evaluation. Section 2.2 identifies key dimensions of scientific critique in peer review. Section 2.3 introduces the theory-guided frameworks used to map these dimensions into measurable indicators.

### 2.1 LLMs as Tools for Scholarly Evaluation

A growing body of work examines LLMs as tools for evaluating scholarly texts. Recent studies suggest that LLMs can perform complex text-evaluation tasks, including research quality assessment, societal impact assessment, and qualitative coding, while emphasizing the need for systematic validation (Kousha & Thelwall, 2025; Thelwall, 2025a, 2025b). LLM-assisted annotation shows a similar tension: LLMs can match or outperform crowd workers on some text-classification tasks, yet may introduce systematic bias in qualitative coding (Ashwin et al., 2025; Gilardi et al., 2023). This literature suggests that LLMs can scale some forms of scholarly evaluation, but their outputs should not be treated as neutral measurements.

The most directly relevant evidence concerns LLM-generated peer-review feedback. Large-scale comparisons show that GPT-4-generated comments can overlap with human reviewer feedback in Nature-family journals and ICLR, sometimes at a level comparable to the overlap between two human reviewers (Liang, Zhang, et al., 2024). Such feedback may also be perceived as useful. Field-experimental evidence from ICLR 2025 further suggests that LLM feedback can improve the specificity and actionability of human reviewers' comments in real review workflows (Thakkar et al., 2026). However, overlap, usefulness, and actionability do not necessarily show that LLMs emphasize the same scientific problems or evaluative priorities as human reviewers.

LLMs are also already present in review practice. Estimates from ICLR 2024 and related AI conferences suggest substantial LLM modification in submitted peer reviews, with higher estimated use in lower-confidence reviews, near-deadline reviews, and reviews from less rebuttal-engaged reviewers (Liang, Izzo, et al., 2024). Detection-oriented studies similarly treat LLM-generated or LLM-modified review text as an emerging governance issue (Rao et al., 2025; Shen & Wang, 2026). These studies establish the practical importance of LLM in peer review, but focus mainly on prevalence, detection, or governance rather than the functional content of LLM critique.

There is some more direct evidence that LLM review performance is dimension specific. Current LLMs appear limited in assessing scientific novelty, even when prompted or specialized for academic evaluation (W. Wu et al., 2026). Multidimensional peer-review benchmarks suggest that LLM systems can perform well on some dimensions, such as structure or constructiveness, but no single system consistently matches a balanced human baseline across review quality, novelty assessment, flaw identification, major-issue prioritization, and depth of analysis (Loc et al., 2026). Work on misinformed review points also shows that weaknesses and questions can contain unsupported premises or ask about issues already addressed in the paper, with LLM judgments showing only moderate agreement with expert labels (Ryu et al., 2026). These studies suggest that LLM reviews are not simply better or worse than human reviews; their performance varies across specific review functions.

Taken together, this literature shows that LLMs can generate useful review-like feedback and participate in peer-review workflows, but it does not systematically explain how scientific critique functions are distributed across human and LLM-generated reviews. Less is known about whether the two differ in their emphasis on high-order weaknesses, argumentative support, mechanistic questioning, integrative reasoning, or revision guidance. This gap motivates the present study's move beyond human-likeness toward a functional profile of LLM scientific critique.

### 2.2 Dimensions of Scientific Critique in Peer Review

Peer review is often described as a quality-control mechanism, but the quality being evaluated is multidimensional. Review-quality instruments reflect this by including recurring targets of reviewer judgment, such as research importance, originality, methodological strengths and weaknesses, presentation, interpretation of results, constructiveness, and substantiation of comments (Jefferson et al., 2002; van Rooyen et al., 1999). Systematic reviews similarly show that review reports are evaluated through relevance, methodological critique, clarity, constructiveness, fairness, thoroughness, and usefulness for authors and editors (Sizo et al., 2026; Superchi et al., 2019). This literature suggests that scientific critique involves more than detecting errors: it requires judgments about what matters scientifically, whether evidence supports claims, and how a manuscript can be improved.

One important dimension of critique is the target of criticism. Empirical analyses show that reviewer comments are distributed across methodology, theory, writing, originality, impact, clarity, soundness, and recommendation rather than concentrated in a single type of problem (Dycke et al., 2023; Kang et al., 2018; Stephen, 2022). Different targets imply different forms of evaluative work. Some comments identify local problems in facts, reporting, or procedures, whereas others challenge conceptual contribution, scientific framing, or limitation awareness. This motivates our distinction between local critique and higher-level critique of conceptual contribution and scientific positioning.

A second dimension is how criticism is justified. A weakness is more useful when it is not merely asserted but supported by evidence, examples, or reasoning. Substantiation has long been treated as a component of review quality, and computational studies show that peer reviews contain argumentative elements, such as evaluations, requests, facts, references, quotations, and claim-evidence relations (Guo et al., 2023; Hua et al., 2019; van Rooyen et al., 1999). Studies of review utility similarly emphasize actionability, grounding, specificity, verifiability, and helpfulness (P. Bharti et al., 2022; P. K. Bharti et al., 2024; Purkayastha et al., 2025; Sadallah et al., 2025). These findings indicate that scientific critique depends not only on what problem is identified, but also on whether the criticism is supported in a way that authors can evaluate and use.

A third dimension is critique through questioning. Reviewer questions are not merely requests for missing information; they can test relationships among claims, evidence, assumptions, and possible revisions. Work on review-rebuttal interaction shows that reviewer comments often initiate exchanges in which authors clarify evidence, defend methodological choices, or revise claims (Cheng et al., 2020; D'Arcy et al., 2024). Scientific question-answering resources built from peer review similarly treat reviewer questions as document-level questions requiring evidence retrieval and answerability judgments (Baumgärtner et al., 2025). Questions are therefore a useful site for examining critique as inquiry: what remains unclear, under-supported, unresolved, or improvable in a paper.

Together, this literature shows that peer review can be analyzed as structured evaluative information rather than as a single score, recommendation, or sentiment judgment. Yet these dimensions have mainly been used to assess review quality, detect problematic reviews, or build computational resources, rather than to compare how critique functions are distributed across human and LLM-generated reviews. This gap motivates our focus on two explicit ICLR review sections: weak points for diagnostic content and argumentative support, and questions for explanatory depth, integrative complexity, and revision-oriented feedback.

### 2.3 Mapping Critique Functions with Theory-Guided Frameworks

Comparing human and LLM critique requires frameworks that translate review text into interpretable dimensions of evaluative function. We use five theory-guided frameworks for this purpose. For weakness critique, Anderson's knowledge types capture the content target of criticism, while Toulmin's argumentation model captures the support structure of criticism. For reviewer questions, Graesser's question-depth framework captures explanatory depth, SOLO taxonomy captures integrative complexity, and Hattie's feedback model captures feedback function. These frameworks are not used to rank review quality, but to identify what reviewers attend to, how they reason, and what kind of author response their comments invite.

The first dimension concerns the knowledge target of critique. Anderson and Krathwohl's revised taxonomy distinguishes factual, conceptual, procedural, and metacognitive knowledge, and was originally developed to classify learning objectives and assessment tasks (Anderson & Krathwohl, 2001; Krathwohl, 2002). Adapted to peer review, factual and procedural weaknesses capture local or operational criticism, such as missing information, implementation details, methods, experiments, or workflow. Conceptual weaknesses capture criticism of theoretical framing, contribution, assumptions, or interpretation. Metacognitive weaknesses capture criticism of self-positioning, limitation awareness, and field-level significance. This distinction allows us to examine whether human and LLM reviewers locate problems at different scientific levels.

A second dimension concerns the argumentative form of weakness critique. Toulmin's model distinguishes claims, data, warrants, backing, qualifiers, and rebuttals, making it useful for analyzing whether criticism is merely asserted or developed through evidence and reasoning (Toulmin, 2003). The model has been widely used in science education to examine how evidence is connected to claims and how arguments address limitations or counter-positions (Erduran et al., 2004). It also aligns with computational studies showing that peer reviews contain argumentative propositions and claim-evidence relations (Fromm et al., 2021; Guo et al., 2023; Hua et al., 2019). In this study, Toulmin's framework distinguishes claim-like weaknesses from evidence-supported, reasoned, contextualized, or dialectical critique.

For reviewer questions, the first relevant dimension is explanatory depth. Question-depth research shows that questions vary not only in frequency but also in cognitive quality: some request surface clarification, whereas others require causal, functional, interpretive, or explanatory reasoning (Graesser et al., 2014; Graesser & Person, 1994). In peer review, this distinction separates questions that ask authors to define a term or report a detail from those that ask why a method works, how evidence supports a claim, or what assumptions and alternative explanations remain. Question depth therefore captures whether reviewer questions function as shallow clarification or deeper scientific inquiry.

The second property of reviewer questions is integrative complexity. The SOLO taxonomy distinguishes responses that address isolated elements from those that integrate multiple elements into a coherent structure or extend them to more abstract contexts (Biggs & Collis, 1989). In peer review, this helps identify whether a question targets a single local issue or requires authors to connect multiple parts of a study, such as motivation, assumptions, methods, evidence, limitations, and implications.

A third property is feedback function. Feedback research distinguishes comments that correct task-level issues from those that address process understanding or guide future improvement (Hattie & Timperley, 2007; Wisniewski et al., 2020). This framework has also been applied to written feedback in higher education, including thesis-related feedback (Lipsch-Wijnen & Dirkx, 2022). Adapted to peer review, task-level questions ask authors to fix or clarify a specific issue, process-oriented questions ask them to explain the reasoning behind a method, result, or claim, and feed-forward questions point toward concrete revision or future improvement.

Together, these frameworks translate weakness points and question points into interpretable indicators of critique function: what is criticized, how criticism is supported, and what kind of inquiry or revision response is invited. This supports the paper's central aim of moving beyond overall human-likeness to examine which functions of scientific critique are reproduced, amplified, under-represented, or reconfigured in LLM-generated peer review.

## 3 Methods

Figure 1 presents the overall methodological framework. The study constructed comparable review-section text from two sources: human weakness and question sections extracted from ICLR reviews, and LLM-generated review text produced from full papers under baseline and expert prompts. These texts were decomposed into weakness and question points and annotated using five theory-guided frameworks: Anderson and Toulmin for weakness critique, and Graesser, SOLO, and Hattie for scientific questions. Annotation quality was assessed through a human audit and cross-model robustness checks that varied the LLM used for review generation and framework annotation. Point-level labels were then aggregated into paper-level metrics to compare human, baseline-prompt LLMs, and expert-prompt LLM critique profiles.

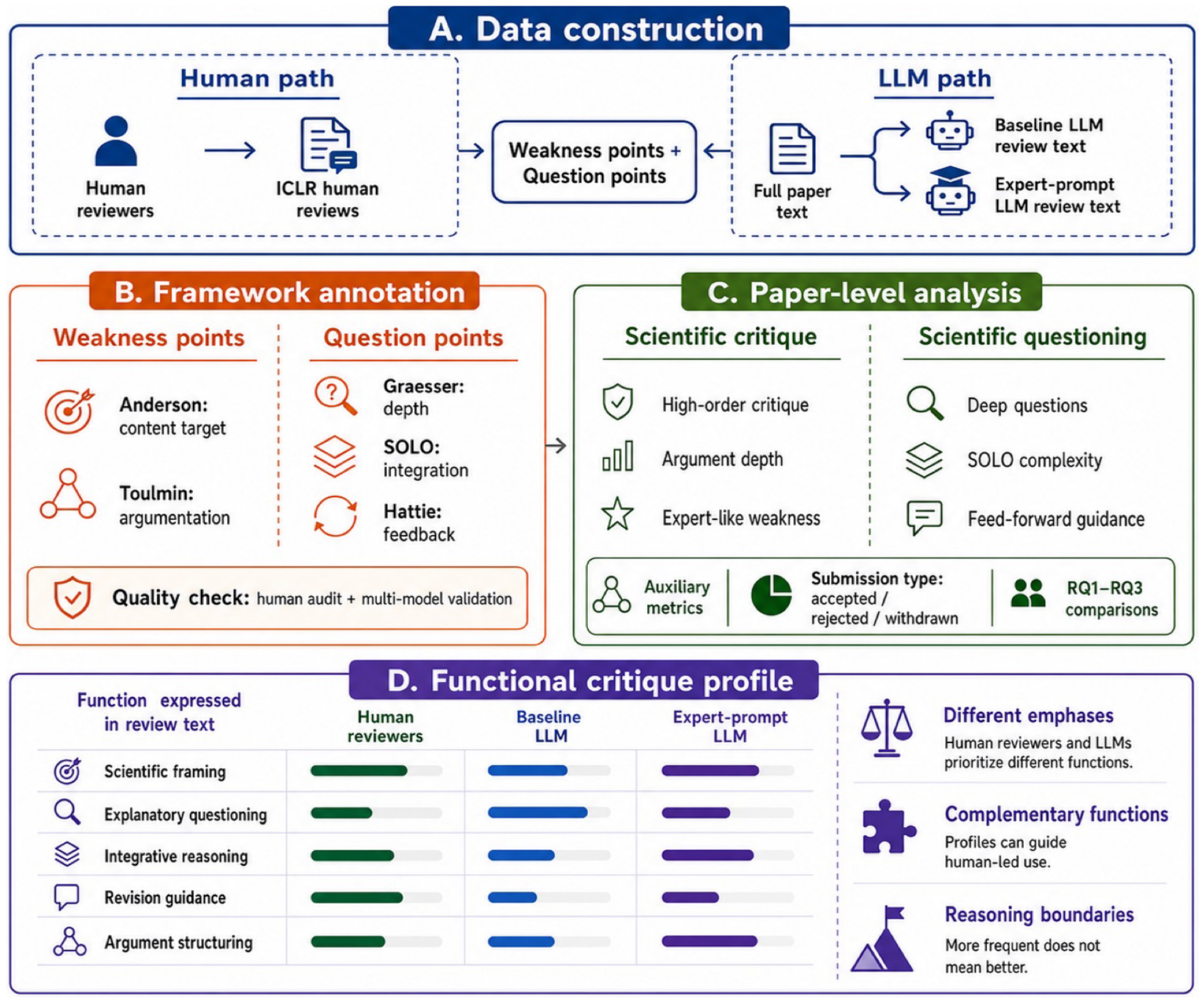

**Figure 1. Methodological framework: Stages A, B and C lead to the output D.**

### 3.1 Data Construction

The empirical setting was ICLR 2025 peer review. ICLR was selected because it is a large, prestigious machine-learning venue with structured open-review reports, standardized review sections, and accessible paper-level metadata. The corpus contained reviewer reports linked to paper identifiers, review identifiers, titles, and submission-type metadata, including accepted, rejected, withdrawn, and a small number of desk-rejected submissions. Submission type was retained to test whether human-LLM differences were stable across outcome groups.

Rather than analyzing full review reports as undifferentiated documents, we focused on two sections that most directly express scientific critique. Weakness sections capture critique as diagnosis: what reviewers identify as problematic and how they justify that judgment. Question sections capture critique as inquiry: what reviewers ask authors to clarify, justify, extend, or revise. This distinction allowed us to compare the content and argumentation of criticism with the depth, integration, and feedback function of scientific questioning.

In the human path, weakness and question sections were extracted from ICLR human reviews for each paper-review pair. In the LLM path, full paper text was used to generate review text under two prompt conditions. The baseline prompt followed standard ICLR-style review instructions, whereas the expert prompt preserved the same output structure but placed greater emphasis on research importance, theoretical contribution, claim-method-evidence alignment, alternative explanations, robustness, and field significance. Full prompts are reported in Appendix B.

Full-text inputs for LLM review generation were obtained from the PDF field of each OpenReview submission record and parsed into section-level text. Thus, the LLM reviews were generated from OpenReview submission PDFs available at the time of data collection, rather than from external arXiv records or separate proceedings files.

All human and LLM review texts were decomposed into point-level weakness and question units. The main human-baseline LLM comparison used the full matched corpus. The expert-prompt analysis used a balanced stratified sample of 600 papers, with 200 accepted, 200 rejected, and 200 withdrawn submissions. This stage was designed as a prompt-sensitivity experiment rather than a second full-corpus estimation. Table 1 summarizes the resulting data structure.

**Table 1. Data construction summary**

| Data component | Paper count | Review / text count | Weakness points | Question points | Notes |
|---|---|---|---|---|---|
| Original ICLR human review corpus | 11,560 | 46,748 | | - | Accepted, rejected, withdrawn, and desk-rejected submissions |
| Human review point-level data | 11,560 | 46,748 | 321,245 | 155,931 | Extracted from human weakness and question sections |
| Baseline-prompt LLM point-level data | 11,560 | 11,560 | 155,738 | 115,149 | Generated from full paper text |
| Expert-prompt stratified sample | 600 | 600 | 4,562 | 4,717 | 200 accepted, 200 rejected, 200 withdrawn |

### 3.2 LLM Configuration and Prompting

LLMs were used in two roles: generating review text from full papers and annotating point-level review text with the five classification frameworks. Separating these roles allowed us to distinguish the object of analysis, LLM-generated scientific critique, from the measurement procedure used to label that critique.

All models were downloaded and run locally on the university high-performance computing cluster rather than accessed through commercial APIs. This provided a consistent execution environment and avoided sending paper or review text to external services.

The main analysis used Gemma3-27B for both review generation and framework annotation. Gemma3-27B was selected because it is a large open-weight model that could be run reproducibly in the same computing environment and has shown good performance in academic evaluation tasks (Thelwall & Mohammadi, 2026). To test model-family sensitivity, we also used Qwen3-32B as an independent reasoning-oriented model. The robustness design crossed the generation and annotation models, producing four combinations: Gemma-generated reviews annotated by Gemma, Gemma-generated reviews annotated by Qwen, Qwen-generated reviews annotated by Gemma, and Qwen-generated reviews annotated by Qwen. The Gemma-Gemma setting was used for the main analysis; the other combinations tested whether the findings depended on the annotation model, the generation model, or both.

### 3.3 Review Point Annotation and Validation

Point-level review texts were annotated using the five theory-guided classification schemes introduced in the literature review. Weakness points were coded for critique content and argumentative structure using Anderson's knowledge types and Toulmin's argumentation model. Question points were coded for explanatory depth, integrative complexity, and feedback function using Graesser's question-depth framework, SOLO taxonomy, and Hattie's feedback model. Table 2 summarizes how these frameworks translated review points into measurable critique functions.

Toulmin elements were coded as separate binary features because data, warrant, backing, qualifier, and rebuttal are not mutually exclusive and could co-occur within a single weakness point. For analysis, we derived an argument category from the strongest support-related element present: claim-only, evidence-supported, reasoned, contextualized, or dialectical critique. This summary category was used to calculate argument-depth metrics, while qualifiers were retained separately as an indicator of uncertainty calibration.

**Table 2. The five theory-guided classification schemes used to classify review points.**

| Text type | Framework | What it captures | Label categories / elements | Main derived metrics |
|---|---|---|---|---|
| Weakness | **Anderson knowledge types** | Content target of critique | Factual; Procedural; Conceptual; Metacognitive | High-order critique rate |
| | **Toulmin argumentation** | Argumentative structure of critique | **Binary elements:** Claim; Data; Warrant; Backing; Qualifier; Rebuttal; **Summary category**: claim-only, evidence-supported, reasoned, contextualized, dialectical | Argument depth; evidence-supported; reasoned critique |
| Question | **Graesser question depth** | Cognitive depth of questions | Shallow; Intermediate; Deep | Deep-question rate; question depth score |
| | **SOLO taxonomy** | Integrative complexity | Pre-structural; Uni-structural; Multi-structural; Relational; Extended abstract | Relational question rate; SOLO complexity score |
| | **Hattie feedback function** | Functional orientation of feedback | Task; Process; Feed-forward | Task/process/feed-forward rates |

All annotations were produced using an LLM-based classifier with framework-specific prompts. Annotation quality was evaluated through a human audit of a stratified subset of 300 weakness and question points. Because the paper-level analyses rely on derived critique indicators, we assessed agreement at the level of research-question-aligned core metrics. Human-Gemma3 agreement was moderate, while Human-Qwen3 agreement was consistently higher across these metrics (Table 3). Full fine-grained framework agreement is reported in Appendix Table A1.

**Table 3. Core Metric Agreement Between Human Audit Labels and LLM Annotations**

| Domain | Core metric | N | Gemma3 agreement | Gemma3 kappa | Qwen3 agreement | Qwen3 kappa |
|---|---|---|---|---|---|---|
| Weakness | High-order critique | 150 | 0.780 | 0.542 | 0.860 | 0.715 |
| | Reasoned-or-above critique | 150 | 0.653 | 0.307 | 0.887 | 0.730 |
| | Expert-like weakness | 150 | 0.800 | 0.462 | 0.867 | 0.701 |
| Question | Deep question | 150 | 0.707 | 0.377 | 0.807 | 0.620 |
| | Integrative question | 150 | 0.780 | 0.576 | 0.893 | 0.785 |
| | Feed-forward guidance | 150 | 0.740 | 0.444 | 0.927 | 0.774 |

### 3.4 Paper-level Metrics

Point-level annotations were aggregated to the paper level. For each paper $p$ and source $s$, where $s$ denotes human review, baseline-prompt LLM review, or expert-prompt LLM review, we computed separate metrics for weakness points and question points. This aggregation allows matched comparisons between human and LLM critique on the same papers.

#### 3.4.1 Weakness Critique Metrics

For weakness critique, we focused on three core metrics. First, **high-order critique rate** measured the proportion of weakness points coded as conceptual or metacognitive:

$$HighOrder_{ps} = \frac{\sum_{i \in W_{ps}} 1(A_i \in \{Conceptual, Metacognitive\})}{|W_{ps}|} \quad (1)$$

Second, **argument depth score** measured the degree to which a weakness was developed as an argument. This score was derived from the binary Toulmin annotations, but it should not be interpreted as a linear ordering of all Toulmin elements. Because qualifiers indicate uncertainty calibration rather than argumentative elaboration, they were excluded from the depth score. Each weakness point received a depth score based on the highest support-related Toulmin element present: rebuttal, backing, warrant, data, or claim-only. The paper-level argument depth score was the mean depth score across all weakness points for paper $p$ and source $s$.

$$D_i = \begin{cases} 5, & if\ Rebuttal_i = 1 \\ 4, & if\ Backing_i = 1 \\ 3, & if\ Warrant_i = 1 \\ 2, & if\ Data_i = 1 \\ 1, & otherwise \end{cases} \quad (2)$$

$$ArgumentDepth_{ps} = \frac{1}{|W_{ps}|} \sum_{i \in W_{ps}} D_i \quad (3)$$

Third, **expert-like weakness rate** measured the proportion of weakness points that were both high-level in content and sufficiently argued. A weakness point was counted as expert-like if it was conceptual or metacognitive under Anderson and reasoned, contextualized, or dialectical under the Toulmin-derived argument category.

$$ExpertLikeWeakness_{ps} = \frac{\sum_{i \in W_{ps}} 1\left[A_i \in \left\{\begin{matrix} Conceptual, \\ Metacognitive \end{matrix}\right\} \wedge T_i \in \left\{\begin{matrix} Reasoned\ critique, \\ Contextualized\ critique, \\ Dialectical\ critique \end{matrix}\right\}\right]}{|W_{ps}|} \quad (4)$$

We also compute auxiliary weakness metrics, including factual, procedural, conceptual, and metacognitive critique rates; evidence-supported rate ($Data = 1$); reasoned critique rate ($Warrant = 1$); contextualized critique rate ($Backing = 1$); dialectical critique rate ($Rebuttal = 1$); and uncertainty calibration rate ($Qualifier = 1$).

### 3.4.2 Scientific Questioning Metrics

For scientific questioning, we focused on three core metrics. **Deep question rate** measured the proportion of questions labeled as deep under Graesser's framework.

$$DeepQuestion_{ps} = \frac{\sum_{j \in Q_{ps}} 1(G_j = Deep)}{|Q_{ps}|} \quad (5)$$

**SOLO complexity score** measured the average integrative complexity of questions by assigning ordinal scores to SOLO categories.

$$SOLOScore_{ps} = \frac{1}{|Q_{ps}|} \sum_{j \in Q_{ps}} S_j \quad (6)$$

where $S_j$ corresponds to the ordered SOLO level of question $j$.

**Feed-forward rate** measured the proportion of questions that guided future revision or improvement under Hattie's feedback model.

$$FeedForward_{ps} = \frac{\sum_{j \in Q_{ps}} 1(H_j = Feed-forward)}{|Q_{ps}|} \quad (7)$$

We also computed auxiliary question metrics, including question depth score, relational question rate, task rate, process rate, mechanistic deep-question rate, and revision-oriented deep-question rate.

### 3.5 Statistical Comparisons and Prompt Effects

To answer RQ1 and RQ2, human and baseline-prompt LLM values were compared at the matched-paper level. For each metric M, we computed the paired difference:

$$\Delta_p^{Baseline-Human} = M_{p,Baseline\ LLM} - M_{p,Human} \tag{8}$$

We reported mean paired differences, matched-pairs rank-biserial effect sizes, and Wilcoxon signed-rank tests. We also repeated the comparison within accepted, rejected, and withdrawn submissions to assess whether the observed human-LLM differences were stable across submission outcomes.

To answer RQ3, we evaluated whether expert prompting moved LLM-generated review text closer to the human review-text profile. On the stratified 600-paper sample, we compared human, baseline-prompt LLM, and expert-prompt LLM values for the same paper-level metrics. For each paper and metric, we computed the absolute human-LLM gap under the baseline and expert prompts:

$$G_p^{Baseline} = \left|M_{p,Baseline\ LLM} - M_{p,Human}\right| \tag{9}$$

$$G_p^{Expert} = \left|M_{p,Expert\ LLM} - M_{p,Human}\right| \tag{10}$$

Prompt-induced convergence was measured as:

$$GapReduction_p = G_p^{Baseline} - G_p^{Expert} \tag{11}$$

A positive value indicated that expert prompting moved the LLM closer to the human profile on that metric; a negative value indicated movement farther away. We also compute the directional prompt shift:

$$\Delta_p^{Expert-Baseline} = M_{p,Expert\ LLM} - M_{p,Baseline\ LLM} \tag{12}$$

This shift identified which critique functions were amplified or reduced by expert prompting.

### 3.6 Prompt Convergence Metrics for Prompt Effects

In addition to the core metrics, we constructed four diagnostic composites from the existing Anderson, Toulmin, Graesser, SOLO, and Hattie labels. These derived metrics were used to distinguish substantive convergence, where expert prompting reduced the human-LLM gap in a critique function, from formal amplification, where it increased expert-like surface features such as explicit support or deeper-sounding questions without moving the LLM closer to the human profile.

For weakness critique, two metrics combined Anderson and Toulmin labels. Supported high-order critique measured the proportion of all weakness points that were conceptual or metacognitive and contained at least one support element: data, warrant, backing, or rebuttal.

$$SupportedHighOrder_{ps} = \frac{\sum_{i \in W_{ps}} 1[A_i \in \{Conceptual, Metacognitive\} \wedge (Data_i \vee Warrant_i \vee Backing_i \vee Rebuttal_i)]}{|W_{ps}|} \tag{13}$$

Support share within high-order critique measured the proportion of high-order weaknesses that contained at least one such support element.

$$SupportShare_{ps} = \frac{\sum_{i \in W_{ps}} 1[A_i \in \{Conceptual, Metacognitive\} \wedge (Data_i \vee Warrant_i \vee Backing_i \vee Rebuttal_i)]}{\sum_{i \in W_{ps}} 1[A_i \in \{Conceptual, Metacognitive\}]} \tag{14}$$

The first metric captured how often supported high-level critique appeared overall; the second asked whether high-level critique was grounded rather than merely abstract.

For questions, two metrics combined Graesser, SOLO, and Hattie labels. Generic deep question rate measured the proportion of questions that were deep under Graesser but not feed-forward under Hattie. This captured deep or mechanistic questioning that did not directly guide revision.

$$GenericDeep_{ps} = \frac{\sum_{j \in Q_{ps}} 1[G_j = Deep \wedge H_j \neq Feed-forward]}{|Q_{ps}|} \quad (15)$$

Substantive revision-oriented deep question rate measured the proportion of questions that were deep, feed-forward, and relational or extended abstract under SOLO. This captured questions that combined explanatory depth, integrative complexity, and revision guidance.

$$SubstantiveDeep_{ps} = \frac{\sum_{j \in Q_{ps}} 1[G_j = Deep \wedge H_j = Feed-forward \wedge S_j \in \{Relational, Extended\ Abstract\}]}{|Q_{ps}|} \quad (16)$$

All four metrics were interpreted relative to the human baseline, not as monotonic quality scores. If expert prompting increased a composite indicator while reducing the human-LLM gap, we interpreted this as substantive convergence. If the indicators increased while the expert-prompt LLM moved farther from the human profile, we interpreted the shift as formal amplification rather than human-like convergence.

## 4 Results

### 4.1 Weakness Critique: Scientific Framing Versus Argumentative Form

The weakness analysis distinguishes what review texts criticized from how the criticism was argued. Human reviews placed greater emphasis on higher-order scientific framing, whereas baseline-prompt Gemma reviews more often developed weaknesses through explicitly Toulmin-style reasoning. This contrast matters because a formally well-argued weakness is not necessarily the same as a scientifically prioritized weakness.

Figure 2 summarizes this contrast using three core indicators. Human reviews had a higher high-order critique rate than baseline-prompt Gemma reviews (Human = 0.451; Gemma = 0.379; $\Delta$ = -0.072, $p$ < 1e-300, matched-pairs rank-biserial effect size $r_{rb}$ = -0.495), indicating that human reviewers more often located weaknesses at the level of conceptual contribution, scientific positioning, broader significance, or limitations. By contrast, baseline-prompt Gemma reviews showed a higher Toulmin argument depth score (Human = 1.581; Gemma = 1.752; $\Delta$ = +0.172, $p$ = 1.40e-189, $r_{rb}$ = +0.330), meaning that their weaknesses more often included explicit reasoning linking an observed issue to an evaluative claim.

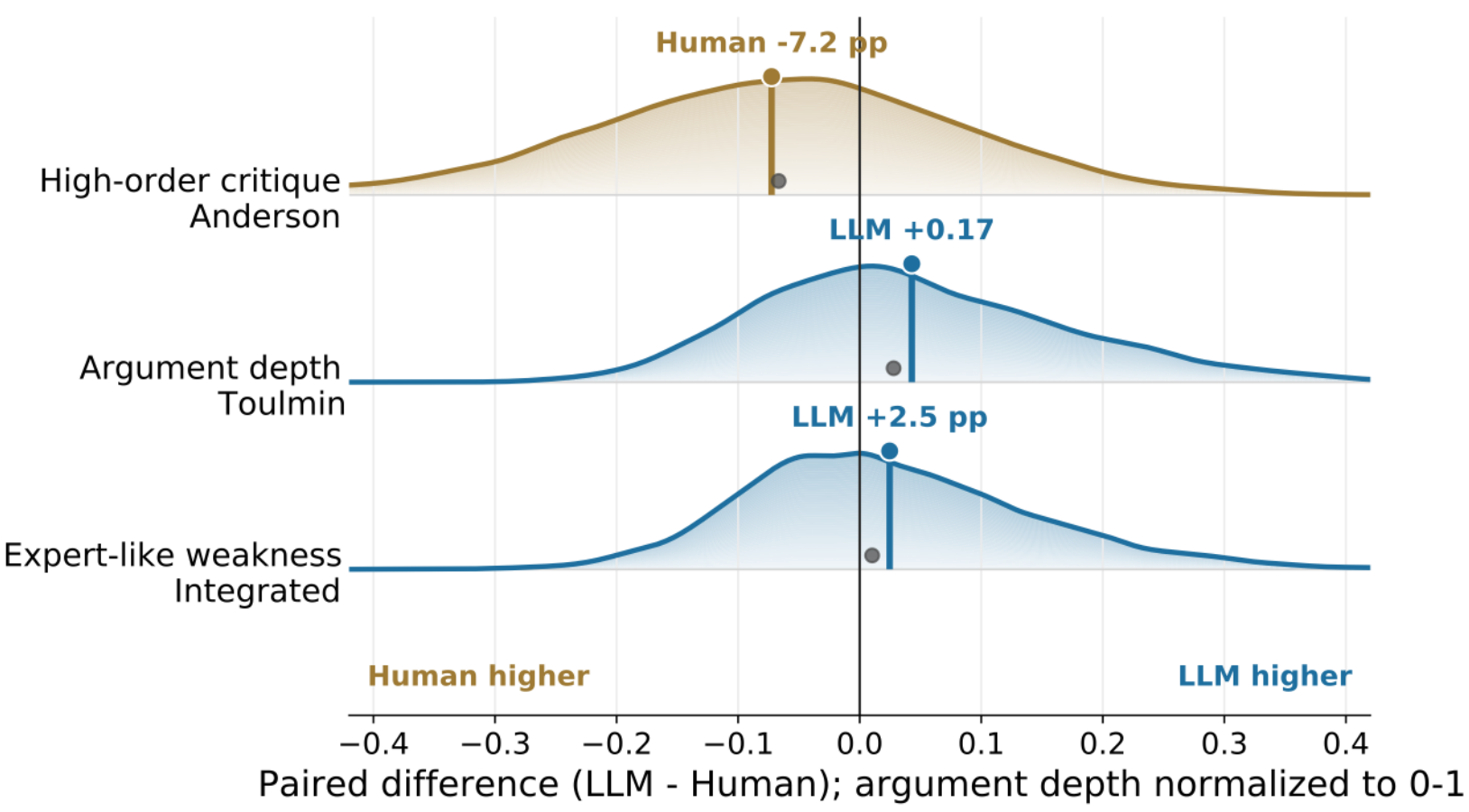

**Figure 2. Core Differences in Weakness Critique Between Human Reviewers and Baseline-Prompt Gemma Reviews.** Values show paper-level paired differences calculated as Baseline Gemma minus Human. Differences for rate-based metrics are expressed in percentage points (pp). Argument

depth was normalized to the 0-1 scale for visualization; the plotted value corresponds to the normalized difference, while the text reports the raw score difference on the original 1-5 scale.

The integrated expert-like weakness metric combines these two dimensions: a weakness must be high-order in content and argumentatively developed. Baseline-prompt Gemma reviews showed a slightly higher value on this metric (Human = 0.091; Gemma = 0.115; $\Delta$ = +0.025, $p$ = 3.23e-62, $r_{rb}$ = +0.190). However, because Gemma reviews showed lower high-order critique rates but higher argument depth, this small difference appears to be driven mainly by argumentative development rather than by greater emphasis on high-order scientific framing.

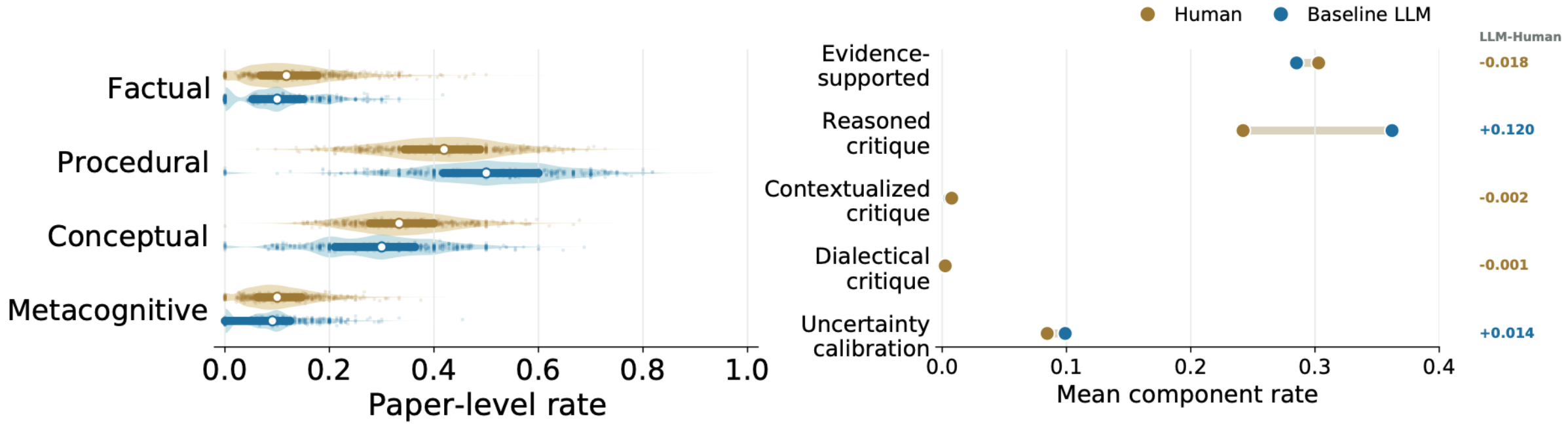

**Figure 3. Content and Argumentation Profiles of Weakness Critique.**

Figure 3 decomposes this pattern. The Anderson distribution shows that Gemma weaknesses were more concentrated in procedural criticism, while human reviewers contained relatively more conceptual and metacognitive critique. The Toulmin decomposition shows that the Gemma advantage in argument depth was driven primarily by reasoned critique, especially warrant-based statements ($\Delta$ = +0.120, $p$ < 1e-300, $r_{rb}$ = +0.474). Differences in contextualized and dialectical critique were small, suggesting that baseline-prompt Gemma reviews more often supplied explicit reasons but not necessarily broader scientific context or engagement with counterarguments. Full auxiliary statistics are reported in Appendix Table A2.

Figure 4 shows that the core weakness differences were stable across accepted, rejected, and withdrawn submissions. Human reviews consistently showed higher high-order critique rates, whereas baseline-prompt Gemma reviews consistently showed higher argument depth and expert-like weakness rates. This suggests that the observed weakness profile was not driven by one submission outcome group.

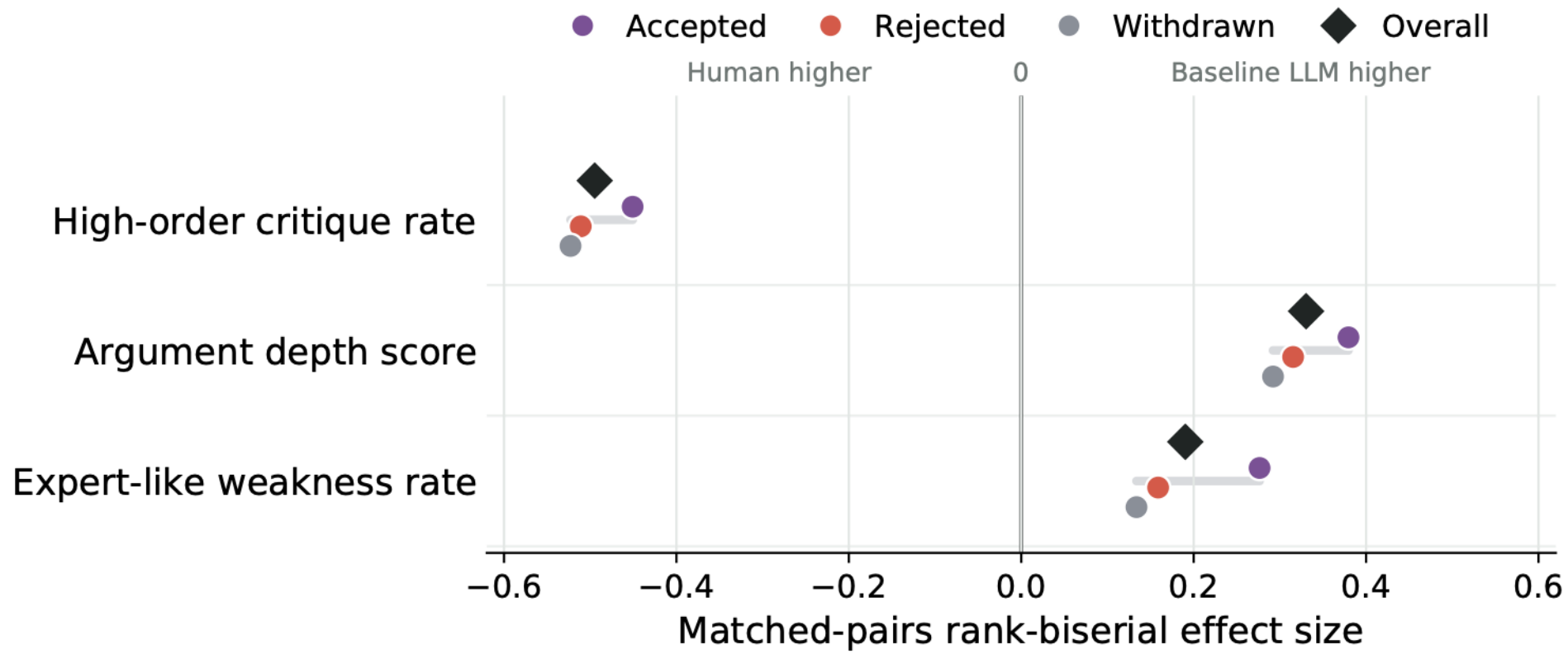

**Figure 4. Human–Baseline Gemma Weakness Differences Across Submission Outcomes**

### 4.2 Scientific Questioning: Mechanistic Depth Versus Revision Guidance

The question analysis shows a different form of human-LLM difference. Whereas the weakness results distinguished critique target from argumentative form, the question results distinguish explanatory ambition from revision guidance. Baseline-prompt Gemma reviews more often contained deep and integrative questions, while human reviews more often used questions as actionable feedback for improvement.

Figure 5 summarizes this pattern across three core frameworks. Under Graesser's question-depth framework, Gemma reviews contained more deep questions than human reviews ($\Delta$ = +0.056, $p$ = 2.51e-183, $r_{rb}$ = +0.318), indicating greater emphasis on causal explanation, mechanism clarification, and justification of assumptions. Under SOLO, Gemma reviews also showed a higher relational-question rate ($\Delta$ = +0.135, $p$ < 1e-300, $r_{rb}$ = +0.722), meaning that their questions more often required authors to connect multiple scientific elements, such as methods, evidence, assumptions, and implications. In contrast, under Hattie's feedback-function framework, human reviews showed a higher feed-forward rate ($\Delta$ = -0.059, $p$ = 1.51e-193, $r_{rb}$ = -0.327), indicating that their questions more often pointed authors toward concrete revisions, improvements, or next steps.

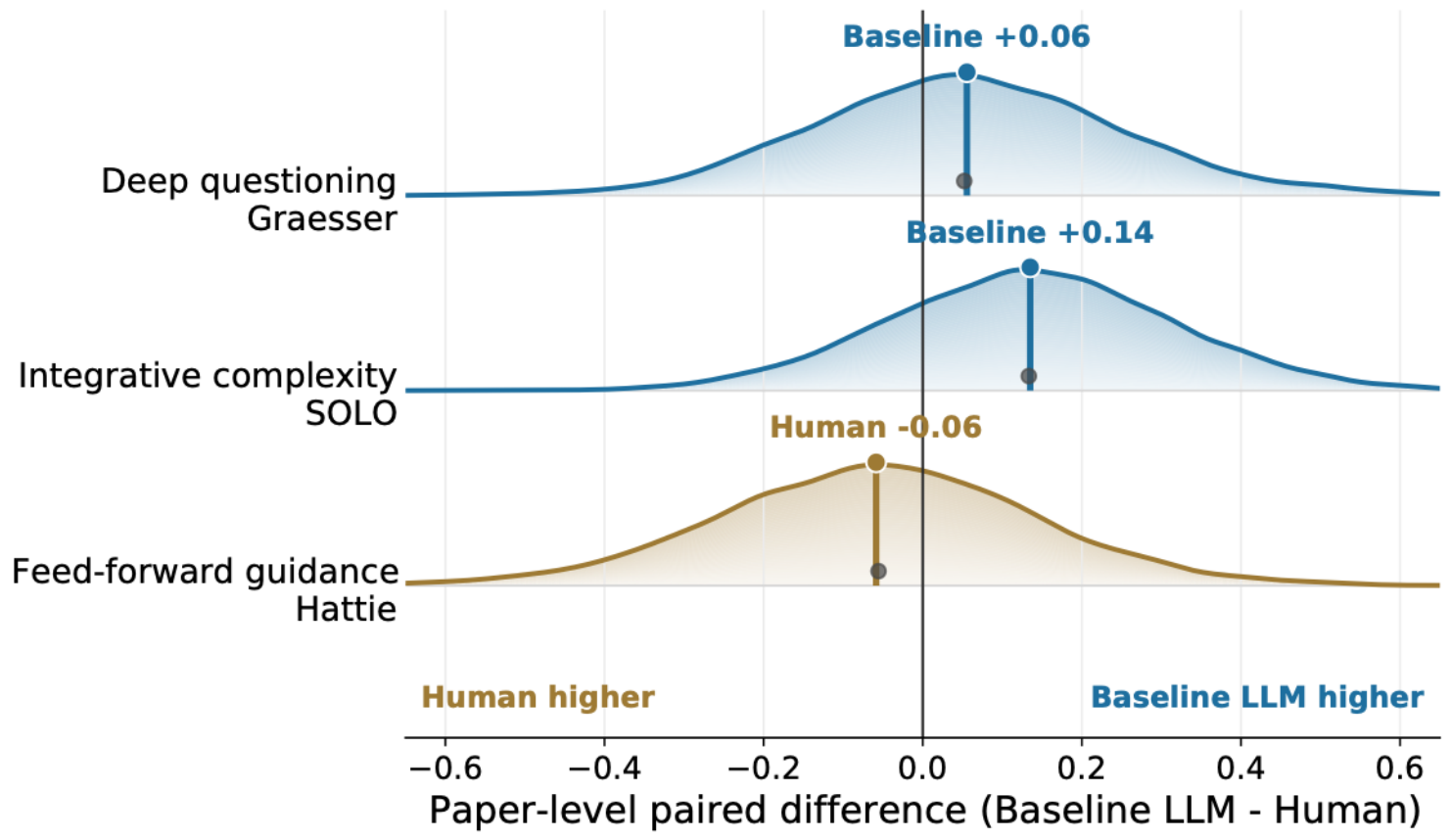

**Figure 5. Core Differences in Scientific Questioning Between Human Reviews and Baseline-Prompt Gemma Reviews**

Figure 6 decomposes this pattern across the three question frameworks. In Hattie's feedback categories, human reviews contained more task-level and feed-forward questions, whereas Gemma reviews contained more process-oriented questions. In Graesser's categories, Gemma reviews contained more deep questions, while human reviews contained more shallow and slightly more intermediate questions. In SOLO, human questions were concentrated in uni-structural forms, whereas Gemma questions showed higher multi-structural, relational, and extended-abstract rates. Full auxiliary statistics are reported in Appendix Table A3.

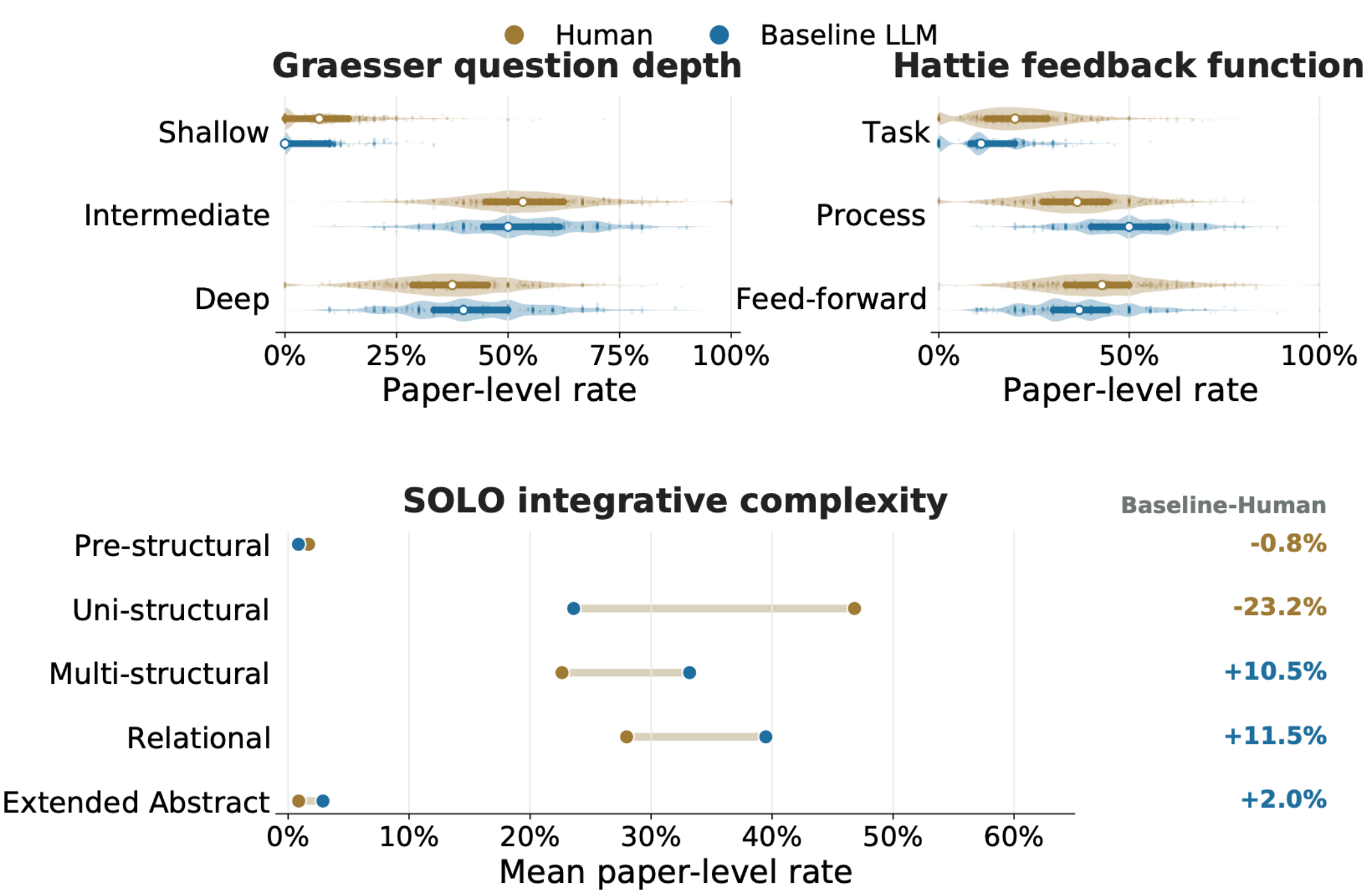

**Figure 6. Question Profiles by Depth, Feedback Function, and Integrative Complexity**

Figure 7 tests whether this pattern depends on submission outcome. The direction of the effects remained stable across accepted, rejected, and withdrawn papers, and the stratified effect sizes closely tracked the overall estimates. This indicates that the differences are not driven by paper quality or decision status but reflect a stable difference in questioning profile. Across both weakness and question metrics, cross-model robustness checks showed that most core human-baseline Gemma directions were reproduced when varying the generation and annotation models. Cross-model robustness result is reported in Appendix Table A5.

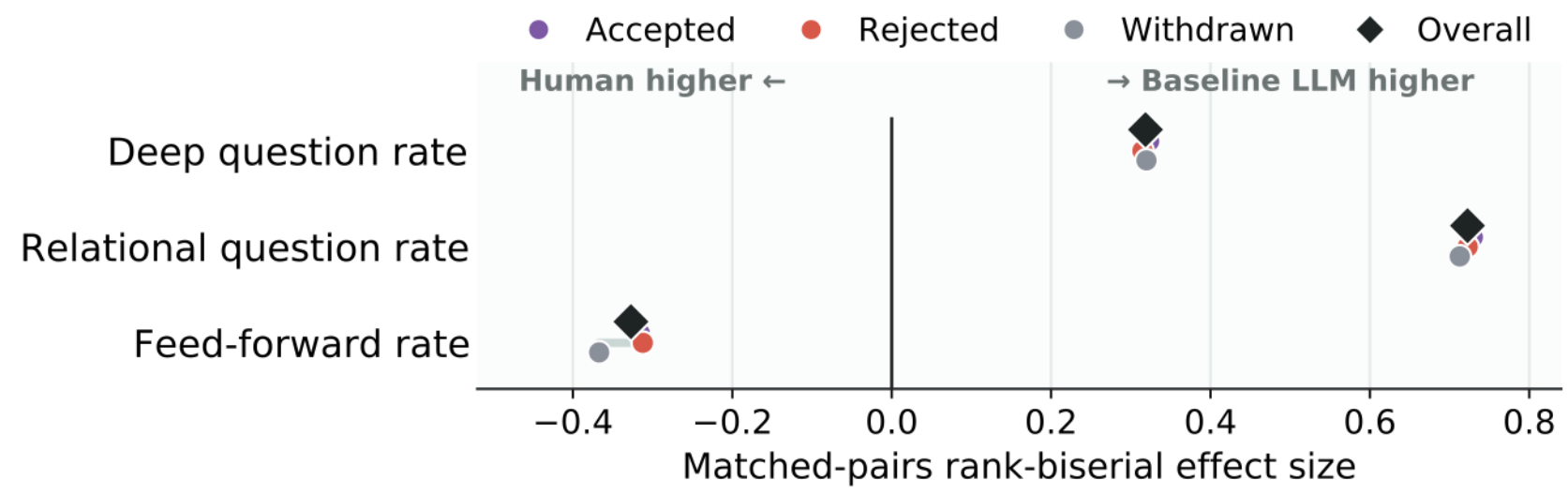

**Figure 7. Human–Baseline Gemma Questioning Differences Across Submission Outcomes**

### 4.3 Expert Prompting: Formal Amplification Rather Than Broad Convergence

The expert-prompt analysis tests whether a more evaluative prompt moved Gemma-generated review text closer to the human review profile. The results show selective change rather than broad convergence: expert prompting shifted several metrics but often amplified the baseline Gemma profile instead of reducing the human-LLM gap.

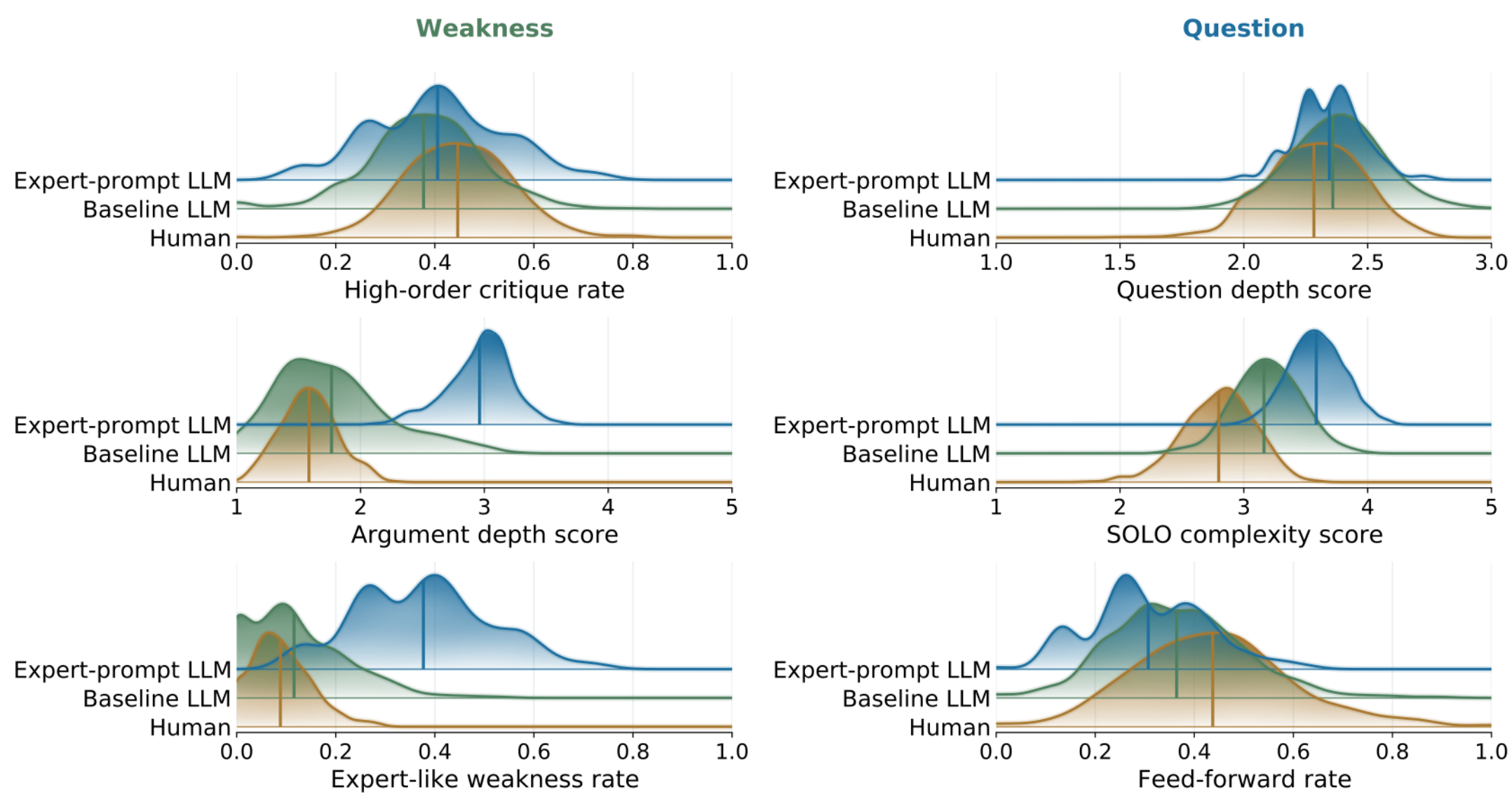

**Figure 8. Expert-Prompt Shifts in Core Weakness and Question Metrics**

Figure 8 shows the shift across six core metrics, with framework-metric convergence statistics reported in Appendix Table A4. For weakness critique, high-order critique moved only modestly toward the human level (Human = 0.446; Baseline = 0.377; Expert = 0.406). The largest shift occurred in argumentative form: the expert-prompt Gemma moved much farther from humans on argument depth (Human = 1.583; Baseline = 1.766; Expert = 2.961) and expert-like weakness rate (Human = 0.088; Baseline = 0.116; Expert = 0.377). Thus, expert prompting mainly increased the formal development of weaknesses rather than reproducing the human distribution of high-level critique.

For questions, expert prompting slightly narrowed the gap in question depth (gap change = +0.047), but moved farther from humans on relational questioning, SOLO complexity, process-level questioning, and feed-forward rate. Feed-forward guidance decreased further below the human level (Human = 0.437; Baseline = 0.364; Expert = 0.307). This indicates that the expert prompt intensified the Gemma tendency toward integrative and process-oriented questioning while weakening alignment with human revision guidance.

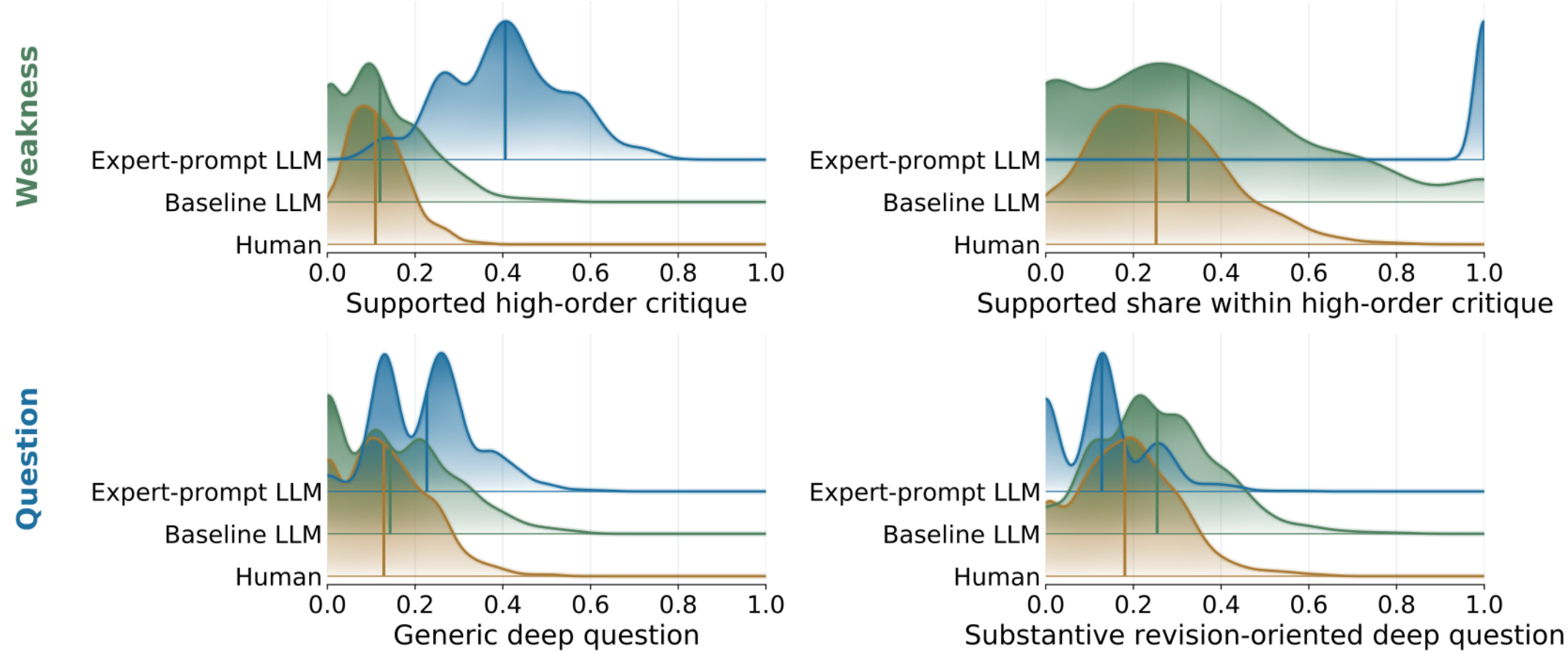

**Figure 9. Substantive Versus Performative Effects of Expert Prompting**

Figure 9 tests this interpretation using four diagnostic composite metrics. For weaknesses, the expert-prompt Gemma strongly increased supported high-order critique and the share of high-order critique with explicit support, overshooting the human distribution. This pattern indicates formalized expert-like critique rather than substantive convergence in critique selection. For questions, expert prompting increased generic deep questioning but did not clearly improve revision-oriented scientific inquiry. Although substantive revision-oriented deep questions moved slightly closer to the human distribution, this reflected a reduction from the baseline-prompt Gemma rather than clear evidence of human-like improvement. Cross-model robustness checks are reported in Appendix Table A6.

| | Human reviewers | Baseline LLM | Expert-prompt LLM | Prompt effect |
|---|---|---|---|---|
| **Scientific framing /value judgment** high-order critique rate | 0.45 | 0.38 | 0.41 | partial convergence |
| **Explanatory questioning** question depth score | 2.28 | 2.36 | 2.35 | partial convergence |
| **Integrative reasoning** SOLO complexity score | 2.80 | 3.17 | 3.59 | amplified LLM profile |
| **Revision guidance** feed-forward rate | 0.44 | 0.37 | 0.31 | less revision guidance |
| **Argument structuring** argument depth score | 1.58 | 1.77 | 2.96 | amplified LLM profile |

**Figure 10. Cognitive Function Profile of Scientific Critique.** Marker size shows row-normalized prominence; numbers are paper-level means over common papers. The figure synthesizes framework-based metrics, not overall review quality; each row is scaled independently.

Figure 10 summarizes the overall functional critique profile in the main Gemma-Gemma analysis. Human reviews placed greater emphasis on scientific framing and revision guidance, whereas baseline-prompt Gemma reviews contained higher rates of explanatory depth, integrative reasoning, and argument structuring. Expert prompting did not produce broad human-like convergence. Although it produced limited movement toward the human profile in scientific framing and explanatory depth, its stronger effect was to amplify the Gemma profile in integrative reasoning and argument structuring, while reducing revision guidance in this main model setting. The central result is therefore not that Gemma-generated reviews are not simply weaker or more human-like versions of human reviews; they represent a different distribution of scientific critique functions.

## 5 Discussion

This study shows that differences between human and LLM-generated peer reviews are not best understood as a single difference in overall review quality, bus as differences in how critique functions are distributed across review text. Its central contribution is to move beyond asking whether LLM reviews are useful or human-like, and instead to map how scientific critique is functionally organized in human and LLM-generated reviews.

This finding refines recent evidence on LLM-generated peer-review feedback. Large-scale comparisons have shown that LLM-generated comments can overlap substantially with human reviewer feedback in both journal and ICLR papers, and that many researchers perceive such feedback as helpful (Liang, Zhang, et al., 2024). Field evidence from ICLR 2025 further shows that an LLM review-feedback agent can influence real reviewers, making revised reviews more specific, actionable, and informative (Thakkar et al., 2026). Our results are consistent with these findings, but show that overlap is not functional equivalence. An LLM comment and a human comment may address a similar issue while differing in what they do within the review: identifying a higher-order problem, supporting a criticism, prompting explanation, connecting evidence, or guiding revision.

The study also extends work on LLM use in AI-conference peer review. Evidence from ICLR and related machine-learning conferences suggests that LLM-modified review text is already present in peer-review ecosystems and may be associated with changes in review style, reviewer behaviour, and textual homogenization (Liang, Izzo, et al., 2024). Detection-oriented studies treat this as a governance issue, asking whether review text has been generated or modified by AI. Our analysis shifts attention from the presence of AI-modified text to its functional signature: what critique functions become more or less visible when LLM-style review text enters peer review.

This interpretation is consistent with ICLR-based and machine-learning peer-review benchmarks that caution against treating LLM reviewers as uniformly reliable. Prior evaluations show that LLMs can perform well on some review-revision dimensions but remain limited in critical feedback, novelty assessment, review-revision reasoning, and balanced multidimensional evaluation (Loc et al., 2026; Xu et al., 2024; Zhou et al., 2024). Our results support this dimension-specific view but add a more interpretable account of where the differences lie: not only whether an LLM receives a high review-quality score, but which critique functions it tends to emphasize.

The human and LLM-generated profiles suggest different emphases in scientific critique. Human reviews more often identified higher-order weaknesses and asked questions directed toward concrete improvement. This aligns with research on peer-review quality, which emphasizes relevance, specificity, constructiveness, thoroughness, and usefulness for authors and editors (Sizo et al., 2026). In our data, human review text more often expressed prioritization: what is scientifically at stake, and what authors should revise.

Baseline-prompt LLM reviews more often contained explanatory questions, higher integrative complexity, and explicitly structured arguments. This suggests that LLMs may function as critique amplifiers: they can broaden the space of possible questions, make warrants explicit, and connect multiple elements of a study. This interpretation is consistent with evidence that AI can be useful in fragmented or combinatorially complex knowledge spaces, where progress depends on connecting dispersed ideas, methods, and evidence (Bianchini et al., 2026), and with human-AI collaboration research showing that complementarity depends on matching task structure to the distinctive contributions of each side (Vaccaro et al., 2024).

The expert-prompt analysis adds an important caution. Strengthening the prompt did not make the LLM uniformly more human-like. It partially narrowed some gaps but mainly intensified the LLM profile: more integrative reasoning and formal argument structuring, alongside less revision guidance. Thus, expert-style language may make LLM critique appear more systematic and authoritative without necessarily improving the underlying evaluative judgment. This is why point-level decomposition and theory-guided functional metrics matter: they make visible which parts of critique are reproduced, amplified, or under-represented, rather than treating fluency, specificity, or apparent expertise as sufficient evidence of review quality.

Practically, the findings support a cautious complementary model of LLM use in scholarly evaluation. For authors, LLMs may be useful as pre-review tools that surface explanatory questions, expose weak method-conclusion links, and make criticism more explicit. For reviewers and editors, LLMs may help broaden the set of issues considered, but human experts remain necessary for judging significance, prioritizing revisions, and making accountable evaluative decisions. For scholarly communication systems, the goal should not be to make LLM reviewers indistinguishable from human reviewers, but to design tools that make their functional profile visible: where they amplify critique, where they over-formalize it, and where human judgment should remain central.

Several limitations should be noted. First, the empirical setting is ICLR 2025, a machine-learning conference review context. ICLR is useful because it provides structured review sections, open-review metadata, and comparability with prior ICLR-based studies, but future work should test whether the same functional profile appears in other disciplines, journals, and review cultures. Second, the main analysis is based on Gemma3-27B-generated reviews, with Qwen3-32B used for robustness checks; additional model families, including domain-specialized and frontier closed models, would strengthen generalizability. Finally, the analysis focuses on the rate at which review points express different functions, lower rates should be interpreted as lower representation in review text, not as proof that LLM cannot produce those functions. Future work should connect these functional metrics to downstream outcomes such as author uptake, revision quality, editorial decisions, and post-publication impact.

## 6 Conclusion

This study examined LLM-generated peer review as a redistribution of evaluative work rather than as a direct approximation of human review. By separating weakness critique from scientific questioning, it showed that human and LLM-generated reviews make different aspects of scientific critique visible.

Human reviews more often foregrounded scientific framing and revision priorities, while LLM-generated reviews more often foregrounded explanation, integration, and explicit argumentation. Expert prompting changed this distribution but did not dissolve it: it made LLM critique more systematic and formally developed, without making it uniformly more human-like.

The implication is that LLMs should not be evaluated only by overall similarity to human reviewers. Their role in peer review depends on the specific critique functions they make more visible, the functions they leave less represented, and the human judgement needed to interpret and govern those differences.

## Reference

Anderson, L. W., & Krathwohl, D. R. (2001). *A taxonomy for learning, teaching, and assessing: A revision of Bloom's taxonomy of educational objectives : complete edition*. Addison Wesley Longman. https://eduq.info/xmlui/handle/11515/18824

Ashwin, J., Chhabra, A., & Rao, V. (2025). Using Large Language Models for Qualitative Analysis can Introduce Serious Bias. *Sociological Methods & Research*, 00491241251338246. https://doi.org/10.1177/00491241251338246

Baumgärtner, T., Briscoe, T., & Gurevych, I. (2025). PeerQA: A Scientific Question Answering Dataset from Peer Reviews. In L. Chiruzzo, A. Ritter, & L. Wang (Eds.), *Proceedings of the 2025 Conference of the Nations of the Americas Chapter of the Association for Computational Linguistics: Human Language Technologies (Volume 1: Long Papers)* (pp. 508–544). Association for Computational Linguistics. https://doi.org/10.18653/v1/2025.naacl-long.22

Bharti, P., Ghosal, T., Agarwal, M., & Ekbal, A. (2022). A Method for Automatically Estimating the Informativeness of Peer Reviews. In Md. S. Akhtar & T. Chakraborty (Eds.), *Proceedings of the 19th International Conference on Natural Language Processing (ICON)* (pp. 280–289). Association for Computational Linguistics. https://aclanthology.org/2022.icon-main.34/

Bharti, P. K., Agarwal, M., & Ekbal, A. (2024). Please be polite to your peers: A multi-task model for assessing the tone and objectivity of critiques of peer review comments. *Scientometrics*, *129*(3), 1377–1413. https://doi.org/10.1007/s11192-024-04938-z

Bianchini, S., Di Girolamo, V., Ravet, J., & Arranz, D. (2026). AI in science: When and where it makes a difference. *Research Policy*, *55*(6), 105478. https://doi.org/10.1016/j.respol.2026.105478

Biggs, J., & Collis, K. (1989). Towards a Model of School-based Curriculum Development and Assessment Using the SOLO Taxonomy. *Australian Journal of Education*, *33*(2), 151–163. https://doi.org/10.1177/168781408903300205

Cheng, L., Bing, L., Yu, Q., Lu, W., & Si, L. (2020). APE: Argument Pair Extraction from Peer Review and Rebuttal via Multi-task Learning. In B. Webber, T. Cohn, Y. He, & Y. Liu (Eds.), *Proceedings of the 2020 Conference on Empirical Methods in Natural Language Processing (EMNLP)* (pp. 7000–7011). Association for Computational Linguistics. https://doi.org/10.18653/v1/2020.emnlp-main.569

D'Arcy, M., Ross, A., Bransom, E., Kuehl, B., Bragg, J., Hope, T., & Downey, D. (2024). ARIES: A Corpus of Scientific Paper Edits Made in Response to Peer Reviews. In L.-W. Ku, A. Martins, & V. Srikumar (Eds.), *Proceedings of the 62nd Annual Meeting of the Association for Computational Linguistics (Volume 1: Long Papers)* (pp. 6985–7001). Association for Computational Linguistics. https://doi.org/10.18653/v1/2024.acl-long.377

Dycke, N., Kuznetsov, I., & Gurevych, I. (2023). NLPeer: A Unified Resource for the Computational Study of Peer Review. In A. Rogers, J. Boyd-Graber, & N. Okazaki (Eds.), *Proceedings of the 61st Annual Meeting of the Association for Computational Linguistics (Volume 1: Long Papers)* (pp. 5049–5073). Association for Computational Linguistics. https://doi.org/10.18653/v1/2023.acl-long.277

Erduran, S., Simon, S., & Osborne, J. (2004). TAPping into argumentation: Developments in the application of Toulmin's Argument Pattern for studying science discourse. *Science Education*, *88*(6), 915–933. https://doi.org/10.1002/sce.20012

Fromm, M., Faerman, E., Berrendorf, M., Bhargava, S., Qi, R., Zhang, Y., Dennert, L., Selle, S., Mao, Y., & Seidl, T. (2021). Argument Mining Driven Analysis of Peer-Reviews. *Proceedings of the AAAI Conference on Artificial Intelligence*, *35*(6), 4758–4766. https://doi.org/10.1609/aaai.v35i6.16607

Gilardi, F., Alizadeh, M., & Kubli, M. (2023). ChatGPT outperforms crowd workers for text-annotation tasks. *Proceedings of the National Academy of Sciences*, *120*(30), e2305016120. https://doi.org/10.1073/pnas.2305016120

Graesser, A. C., Li, H., & Forsyth, C. (2014). Learning by Communicating in Natural Language With Conversational Agents. *Current Directions in Psychological Science*, *23*(5), 374–380. https://doi.org/10.1177/0963721414540680

Graesser, A. C., & Person, N. K. (1994). Question Asking During Tutoring. *American Educational Research Journal*, *31*(1), 104–137. https://doi.org/10.3102/00028312031001104

Guo, Y., Shang, G., Rennard, V., Vazirgiannis, M., & Clavel, C. (2023). Automatic Analysis of Substantiation in Scientific Peer Reviews. In H. Bouamor, J. Pino, & K. Bali (Eds.), *Findings of the Association for Computational Linguistics: EMNLP 2023* (pp. 10198–10216). Association for Computational Linguistics. https://doi.org/10.18653/v1/2023.findings-emnlp.684

Hattie, J., & Timperley, H. (2007). The Power of Feedback. *Review of Educational Research*, *77*(1), 81–112. https://doi.org/10.3102/003465430298487

Hua, X., Nikolov, M., Badugu, N., & Wang, L. (2019). Argument Mining for Understanding Peer Reviews. In J. Burstein, C. Doran, & T. Solorio (Eds.), *Proceedings of the 2019 Conference of the North American Chapter of the Association for Computational Linguistics: Human Language Technologies, Volume 1 (Long and Short Papers)* (pp.

2131–2137). Association for Computational Linguistics. https://doi.org/10.18653/v1/N19-1219

Jefferson, T., Wager, E., & Davidoff, F. (2002). Measuring the Quality of Editorial Peer Review. *JAMA*, *287*(21), 2786–2790. https://doi.org/10.1001/jama.287.21.2786

Kang, D., Ammar, W., Dalvi, B., van Zuylen, M., Kohlmeier, S., Hovy, E., & Schwartz, R. (2018). A Dataset of Peer Reviews (PeerRead): Collection, Insights and NLP Applications. In M. Walker, H. Ji, & A. Stent (Eds.), *Proceedings of the 2018 Conference of the North American Chapter of the Association for Computational Linguistics: Human Language Technologies, Volume 1 (Long Papers)* (pp. 1647–1661). Association for Computational Linguistics. https://doi.org/10.18653/v1/N18-1149

Kousha, K., & Thelwall, M. (2025). Assessing the societal influence of academic research with ChatGPT: Impact case study evaluations. *Journal of the Association for Information Science and Technology*, *76*(10), 1357–1373. https://doi.org/10.1002/asi.25021

Krathwohl, D. R. (2002). A Revision of Bloom's Taxonomy: An Overview. *Theory Into Practice*, *41*(4), 212–218. https://doi.org/10.1207/s15430421tip4104_2

Liang, W., Izzo, Z., Zhang, Y., Lepp, H., Cao, H., Zhao, X., Chen, L., Ye, H., Liu, S., Huang, Z., McFarland, D., & Zou, J. Y. (2024, June 6). *Monitoring AI-Modified Content at Scale: A Case Study on the Impact of ChatGPT on AI Conference Peer Reviews*. Forty-first International Conference on Machine Learning. https://openreview.net/forum?id=bX3J7ho18S

Liang, W., Zhang, Y., Cao, H., Wang, B., Ding, D. Y., Yang, X., Vodrahalli, K., He, S., Smith, D. S., Yin, Y., McFarland, D. A., & Zou, J. (2024). Can Large Language Models Provide Useful Feedback on Research Papers? A Large-Scale Empirical Analysis. *NEJM AI*, *1*(8), AIoa2400196. https://doi.org/10.1056/AIoa2400196

Lipsch-Wijnen, I., & Dirkx, K. (2022). A case study of the use of the Hattie and Timperley feedback model on written feedback in thesis examination in higher education. *Cogent Education*, *9*(1), 2082089. https://doi.org/10.1080/2331186X.2022.2082089

Loc, N. P. P., La Viet, T. H., Khanh, T. T., Nguyen, D. A., Pham, T. A. N., Nguyen, T., Chawla, N. V., Buntine, W., Wong, K.-S., Doan, K. D., & Nguyen, B. T. (2026, May 26). *PRISM: A Multi-Dimensional Benchmark for Evaluating LLM Peer Reviewers*. arXiv.Org. https://arxiv.org/abs/2605.26730v2

Lund, B. D., Wang, T., Mannuru, N. R., Nie, B., Shimray, S., & Wang, Z. (2023). ChatGPT and a new academic reality: Artificial Intelligence-written research papers and the ethics of the large language models in scholarly publishing. *Journal of the Association for Information Science and Technology*, *74*(5), 570–581. https://doi.org/10.1002/asi.24750

Mahony, S. (2022). Toward openness and transparency to better facilitate knowledge creation. *Journal of the Association for Information Science and Technology*, *73*(10), 1474–1488. https://doi.org/10.1002/asi.24652

Purkayastha, S., Li, Z., Lauscher, A., Qu, L., & Gurevych, I. (2025). LazyReview: A Dataset for Uncovering Lazy Thinking in NLP Peer Reviews. In W. Che, J. Nabende, E. Shutova, & M. T. Pilehvar (Eds.), *Proceedings of the 63rd Annual Meeting of the Association for Computational Linguistics (Volume 1: Long Papers)* (pp. 3280–3308). Association for Computational Linguistics. https://doi.org/10.18653/v1/2025.acl-long.165

Rao, V. S., Kumar, A., Lakkaraju, H., & Shah, N. B. (2025). Detecting LLM-generated peer reviews. *PLOS ONE*, *20*(9), e0331871. https://doi.org/10.1371/journal.pone.0331871

Ryu, H., Jang, D., Lee, H. S., Jeong, J., Kim, G., Cho, D., Chu, G., Hwang, M., Jang, H., Kim, C., Kim, H., Kim, J., Kim, J., Kim, Y., Lee, K., Park, C., Yun, H., Betz, G., & Yang, E. (2026). *ReviewScore: Misinformed Peer Review Detection with Large Language Models* (arXiv:2509.21679). arXiv. https://doi.org/10.48550/arXiv.2509.21679

Sadallah, A., Baumgärtner, T., Gurevych, I., & Briscoe, T. (2025). The Good, the Bad and the Constructive: Automatically Measuring Peer Review's Utility for Authors. In C. Christodoulopoulos, T. Chakraborty, C. Rose, & V. Peng (Eds.), *Proceedings of the 2025 Conference on Empirical Methods in Natural Language Processing* (pp. 28991–29021). Association for Computational Linguistics. https://doi.org/10.18653/v1/2025.emnlp-main.1476

Shen, S., & Wang, K. (2026). *Detecting AI-Generated Content in Academic Peer Reviews* (arXiv:2602.00319). arXiv. https://doi.org/10.48550/arXiv.2602.00319

Sizo, A., Lino, A., Rocha, Á., & Reis, L. P. (2026). Determining quality dimensions for peer review reports using a Delphi approach. *Scientometrics*, *131*(4), 2133–2183. https://doi.org/10.1007/s11192-026-05603-3

Stephen, D. (2022). Peer reviewers equally critique theory, method, and writing, with limited effect on the final content of accepted manuscripts. *Scientometrics*, *127*(6), 3413–3435. https://doi.org/10.1007/s11192-022-04357-y

Superchi, C., González, J. A., Solà, I., Cobo, E., Hren, D., & Boutron, I. (2019). Tools used to assess the quality of peer review reports: A methodological systematic review. *BMC Medical Research Methodology*, *19*(1), 48. https://doi.org/10.1186/s12874-019-0688-x

Thakkar, N., Yuksekgonul, M., Silberg, J., Garg, A., Peng, N., Sha, F., Yu, R., Vondrick, C., & Zou, J. (2026). A large-scale randomized study of large language model feedback in peer review. *Nature Machine Intelligence*, *8*(3), 326–336. https://doi.org/10.1038/s42256-026-01188-x

Thelwall, M. (2025a). ChatGPT for complex text evaluation tasks. *Journal of the Association for Information Science and Technology*, *76*(4), 645–648. https://doi.org/10.1002/asi.24966

Thelwall, M. (2025b). Evaluating research quality with Large Language Models: An analysis of ChatGPT's effectiveness with different settings and inputs. *Journal of Data and Information Science*, *10*(1), 7–25. https://doi.org/10.2478/jdis-2025-0011

Thelwall, M., & Mohammadi, E. (2026). Can small and reasoning large language models score journal articles for research quality and do averaging and few-shot help? *Scientometrics*. https://doi.org/10.1007/s11192-026-05585-2

Toulmin, S. E. (2003). *The Uses of Argument* (2nd ed.). Cambridge University Press. https://doi.org/10.1017/CBO9780511840005

Vaccaro, M., Almaatouq, A., & Malone, T. (2024). When combinations of humans and AI are useful: A systematic review and meta-analysis. *Nature Human Behaviour*, *8*(12), 2293–2303. https://doi.org/10.1038/s41562-024-02024-1

van Rooyen, S., Black, N., & Godlee, F. (1999). Development of the review quality instrument (RQI) for assessing peer reviews of manuscripts. *Journal of Clinical Epidemiology*, *52*(7), 625–629. https://doi.org/10.1016/s0895-4356(99)00047-5

Wisniewski, B., Zierer, K., & Hattie, J. (2020). The Power of Feedback Revisited: A Meta-Analysis of Educational Feedback Research. *Frontiers in Psychology*, *10*. https://doi.org/10.3389/fpsyg.2019.03087

Wu, K., Wu, E., Wei, K., Zhang, A., Casasola, A., Nguyen, T., Riantawan, S., Shi, P., Ho, D., & Zou, J. (2025). An automated framework for assessing how well LLMs cite relevant medical references. *Nature Communications*, *16*(1), 3615. https://doi.org/10.1038/s41467-025-58551-6

Wu, W., Zhao, Y., Wang, Y., Li, S., Shao, J., Long, Y., & Zhang, C. (2026, April 13). *NovBench: Evaluating Large Language Models on Academic Paper Novelty Assessment*. arXiv.Org. https://arxiv.org/abs/2604.11543v1

Xu, S., Lu, Y., Schoenebeck, G., & Kong, Y. (2024, October 4). *Benchmarking LLMs' Judgments with No Gold Standard*. The Thirteenth International Conference on Learning Representations. https://openreview.net/forum?id=uE84MGbKD7

Zhou, R., Chen, L., & Yu, K. (2024). Is LLM a Reliable Reviewer? A Comprehensive Evaluation of LLM on Automatic Paper Reviewing Tasks. In N. Calzolari, M.-Y. Kan, V. Hoste, A. Lenci, S. Sakti, & N. Xue (Eds.), *Proceedings of the 2024 Joint International Conference on Computational Linguistics, Language Resources and Evaluation (LREC-COLING 2024)* (pp. 9340–9351). ELRA and ICCL. https://aclanthology.org/2024.lrec-main.816/